\documentclass[namedreferences]{solarphysics}

\usepackage[hyperref,optionalrh,showbiblabels]{spr-sola-addons} % For Solar Physics 
\usepackage{graphicx}        % For eps figures, newer & more powerfull
\usepackage{color}           % For color text: \color command
\renewcommand{\vec}[1]{{\mathbfit #1}}

\chardef\us=`\_

\usepackage{amsmath}
\usepackage{natbib}
\usepackage{color}
\ifx\doiurl    \undefined \def\doiurl#1{\href{https://doi.org/#1}{\textsf{DOI}}}\fi
\ifx \urlurl    \undefined \def
	\urlurl#1{\href{http://#1}{\textsf{#1}}}\fi

\begin{document}
\begin{article}
\begin{opening}
\title{Time-Dependent Two-Fluid Magnetohydrodynamic Model and Simulation of the chromosphere}

\author[addressref=aff2,corref,email={qusai.alshidi@gmail.com}]{\inits{Q.}\fnm{Qusai}~\lnm{Al Shidi}\orcid{0000-0003-0426-038X}}%\sep
\author[addressref=aff1,corref]{\inits{O.}\fnm{Ofer}~\lnm{Cohen}\orcid{0000-0003-3721-0215}}%\sep
\author[addressref=aff1]{\inits{P.}\fnm{Paul}~\lnm{Song}}%\sep
\author[addressref=aff1]{\inits{J.-N.}\fnm{Jiannan}~\lnm{Tu}}%\sep
\address[id=aff2]{Climate and Space Science and Engineering, University of Michigan, 2455 Hayward St., Ann Arbor, MI 48109, USA}
\address[id=aff1]{Lowell Center for Space Science and Technology, University of Massachusetts Lowell, 600 Suffolk St., Lowell, MA 01854, USA}

\runningauthor{Q. Al Shidi \textit{et al.}}
\runningtitle{Time-Dependent MHD Model of the chromosphere}

\begin{abstract}
	The Sun's chromosphere is a highly dynamic, partially ionized region where spicules (hot jets of plasma) form. Here we present a two-fluid MHD model to study the chromosphere, which includes ion--neutral interaction and frictional-heating. Our simulation recovers a magnetic-canopy shape that forms quickly, but it is also quickly disrupted by the formation of a jet. Our simulation produces a shock self-consistently, where the jet is driven by the frictional-heating, which is much greater than the ohmic-heating. Thus, our simulation demonstrates that the jet could be driven purely by thermal effects due to ion--neutral collisions and not by magnetic reconnection. We plan to improve the model to include photo-chemical effects and radiation. 
\end{abstract}
\end{opening}

\section{Introduction} 

The base of the outer solar atmosphere (right above the photosphere), referred to as the chromosphere, is highly dynamic and is strongly influenced by magnetic-field activity~\citep{Hasan08}. It is partially ionized, with as much as a thousand times more neutral hydrogen than ions \citep{Alissandrakis2018}. \citet{Song2011} and \citet{Leake2014} discussed in detail the effects of ion--neutral coupling in  the chromosphere by considering neutrals and collisions into the momentum and Maxwell equations.

A local-scale configuration commonly used in chromosphere models are so-called supergranules, convection cells of 30\,Mm in the horizontal directions. Supergranules are separated by strong magnetic-field regions, referred to as a ``network". It is believed that the network regions correspond to strong magnetic-field at the photosphere. Because the thermal-pressure decreases with height while the total magnetic flux is conserved, the magnetic-field tends to expand in higher altitudes to form a canopy or ``wine glass" shaped magnetic geometry so that total pressure, the  magnetic pressure, $p_\text{B}$, plus thermal-pressure, $p_\text{T}$, \textit{i.e.} $p_\text{T}+p_\text{B}$, is balanced at each height \citep{gabriel1976}. This geometry is reflected in a significant altitude change in the plasma beta, which is the ratio of the thermal to magnetic pressures, $\beta=p_\text{T}/p_\text{B}$. The chromosphere has much lower temperature than that of the hot corona. The average temperature of the chromosphere is around 6000\,K while the coronal temperature is over million degrees \citep{AvrettLoeser2008}. This dramatic change in the temperature, taking place in a very narrow region of a few hundred kilometers referred to as the transition region, is known as the coronal heating problem and it is a long-standing problem in solar physics.

In the chromosphere, shocks are often observed and jets are commonplace in the form of spicules. Spicules are local, hot, upward jets of plasma observed in the corona. The origination and mechanism behind spicule formation are not well known. Spicules are categorized into two types \citep{Pontieu2007}. Type I spicules have longer lifetimes and are typically larger (five minutes and $\approx$15\,Mm). Type II spicules have shorter lifetimes and are smaller (less than two minutes and $\approx$6\,Mm). \citet{Pontieu2004} suggested that photospheric oscillations give birth to spicules. It was also suggested that magnetic reconnection may be the cause of Type II spicules \citep{Pontieu2007}.

There is strong observational evidence for the ubiquity of Alfv\'en waves in the Sun's atmosphere \citep{Aschwanden1999,Nakariakov1999,Ofman2008,Jess2009,McIntosh2011,Mathioudakis2013}. These observations of the solar limbs show Alfv\'en waves with amplitudes in the range of 5\,--\,25\,km\,s$^{-1}$ and phase speeds 250\,--\,500\,km\,s$^{-1}$.  \citet{Pontieu2007} have shown that Alfv\'en waves are carried by spicules with significant amplitudes of order 20\,km\,s$^{-1}$. It has been suggested that Alfv\'en waves could be damped in the chromosphere due to collisions between plasma ions and neutrals and cause sufficient heating \citep{Song2011} and \citep{Song2014}. \citet{Song2017} further proposed that the nonuniform heating associated with the Alfv\'en waves may explain the formation of the spicules. \citet{BradyArber2016} studied heating due to Alfv\'en and kink waves turning into shocks and dissipating while still providing sufficient heating albeit a different mechanism. \citet{Khomenko2017} looks at other waves as well as Alfv\'en waves and their effects in weakly ionized solar plasma. 

Recent simulations and models of the chromosphere have adopted a single-fluid approach~\citep{Gudiksen2011,BradyArber2016,Ni2016,Kuzma2017}. In order to account for neutrals' effect on the plasma and electric field an ``ambipolar diffusion" term is added to the Generalized Ohm's Law to account for collisions~\citep{Leake2006,Arber2007,Martinez-Sykora2012,Martinez-Sykora2017}. This is useful to study the highly magnetized, low-$\beta$, environment of the upper chromosphere. A two-fluid approach takes into account separate fluid motions of plasma and neutrals especially in highly collisional environments like the low solar atmosphere in regions of high magnetic gradients. \citet{Soler2013} studied Alfv\'en-wave propagation in partially ionized plasma while taking a generalized approach with free parameters with the lower solar atmosphere case as an example. \citet{Maneva2017} and \citet{Wojcik2018} studied\ acoustic-wave propagation using a collisional two-fluid approach as well. Three-fluid simulations, with electrons being the third fluid, were carried out to study reconnection events in the chromosphere~\citep{Leake2013}. In these highly localized small regions, electron motion is separate from ion and neutral motion and its effects cannot be neglected. The {\sf Bifrost} \citep{Gudiksen2011} code employs a very large artificial viscosity for stability and uses the one-fluid approach. By self-consistently solving plasma, neutral fluids, and Maxwell's equations, \citet{Tu2013} have shown that due to Joule dissipation Alfv\'en waves can provide strong enough chromospheric-heating and that the heating rate is consistent with the theoretical model of \citet{Song2011}. Alternatively, it was also shown that the coupling of Alfv\'en waves with magneto-acoustic waves leads to shock heating that causes sufficient chromospheric-heating \citep{BradyArber2016}.

The aim for this article is to introduce a two-dimensional, self-consistent, and conservative magnetohydrodynamic (MHD) model for the chromosphere that includes heating. In particular, the model accounts for the neutrals as an additional fluid with its interaction to the ions as advocated by \citet{Song2011}. The heating presented in this model is collisional in nature \citep{Vasyliunas2005}, and it is separated into two terms: an ohmic term due to electron collisions and frictional term due to ion--neutral collisions. This model also shows a spontaneous eruptive event due to heating in the magnetic flux concentrations of the solar magnetic network. We show that the spontaneous eruptive event can occur due to the frictional-heating, while magnetic reconnection is not a dominant factor or a requirement.

In Section~\ref{sec:model} and Section~\ref{sec:domain} the details of the model and code are explained. The validation of the code is carried out in Section~\ref{sec:validation}. It is then followed by Section~\ref{sec:results} where expected behavior from the model, such as formation of the magnetic canopy is shown and the eruption is thoroughly investigated.

\section{An MHD Model for the chromosphere}\label{sec:model}

\subsection{Governing Equations}

In the model presented here, a two-fluid, collisional approach is used to account for the friction between ions and neutrals. The first fluid, the ion fluid, is similar to the traditional, single-fluid plasma, and assumes charge quasi-neutrality. With Maxwell's equations, this fluid is used to calculate self-consistently the evolution of the plasma motion, magnetic fields, and currents in the system. The second, neutral-fluid, accounts for the neutral particles in the system. The two fluids interact by collisions. The friction between ions and neutrals affects the neutral fluid motion so that the neutrals experience the electromagnetic effects. 

This is similar to the two-fluid \textsf{HiFi} code used for studying reconnection in the chromosphere \citep{Leake2012} but without the electron pressure. Electrons are not treated as a separated fluid in this model because the electron's mass is more than three orders of magnitude smaller than ion or neutral particle mass, so its contribution to heating and the equations, considering the scales, is negligible. The effects of its motion are included in the current terms. We use empirical number-density model that comes from observations assuming steady-state in the chromosphere~\citep{AvrettLoeser2008}. Here, we do not account for ionization and recombination which we plan to implement photo-chemistry in future work.

Our model uses Cartesian, 2.5 dimensionality, which means that all the quantities are functions of horizontal [$x$] and vertical [$z$] directions but uniform in the $y$-direction, while vector quantities  have all three components including the third component that is perpendicular [$\perp$] to the $x$--$z$-plane. The perpendicular components are used to generate Alfv\'en waves at the photospheric boundary. 

The governing equations are adapted from \citet{Tu2016} for the two-fluid (plasma and neutral) system as follows:

\begin{equation}
	\frac{\partial \rho_\text{i}}{\partial t} + \nabla \cdot \rho_\text{i} \vec{V} = 0
\end{equation}

\begin{equation}
	\frac{\partial \rho_\text{n}}{\partial t} + \nabla \cdot \rho_\text{n} \vec{U} = 0
\end{equation}

\begin{equation}\label{eq:plasmamomentum}
	\frac{\partial \rho_\text{i} \vec{V}}{\partial t} = - \nabla \cdot [\rho_\text{i} \vec{V}\vec{V} + n_\text{i} \text{k}_\text{b} T_\text{i} + \frac{B^2}{2 \mu_0}\vec{I} - \frac{\vec{BB}}{\mu_0}] + \rho_\text{i} \vec{g} - \rho_\text{i} \nu_\text{in} (\vec{V} - \vec{U})
\end{equation}

\begin{equation}\label{eq:neutralmomentum}
	\frac{\partial \rho_\text{n} \vec{U}}{\partial t} = - \nabla \cdot [\rho_\text{n} \vec{U}\vec{U} + n_\text{n} \text{k}_\text{b} T_\text{n}] + \rho_\text{n} \vec{g} + \rho_\text{i} \nu_\text{in} (\vec{V} - \vec{U})
\end{equation}

\begin{equation}
	\frac{\partial \vec{B}}{\partial t} = - \nabla \cdot ( \vec{V}\vec{B} - \vec{B}\vec{V} ) - \nabla \times ( \eta \vec{J}  + \frac{1}{n_\text{e} e} \vec{J} \times \vec{B} )
\end{equation}

\begin{equation}\label{eq:energyion}
\begin{aligned}
	\frac{\partial \text{k}_\text{b} T_\text{i}}{\partial t} = - \nabla \cdot (\text{k}_\text{b} T_\text{i} \vec{V}) + \frac{1}{3} \text{k}_\text{b} T_\text{i} \nabla \cdot \vec{V} - \frac{2}{3} \frac{\nabla \cdot (\vec{q}_\text{i} + \eta \vec{B} \times \vec{J})}{n_\text{i}} \\
	+ \frac{2}{3} \frac{\text{m}_\text{i}}{\text{m}_\text{i}+\text{m}_\text{n}} \nu_\text{in} [3 \text{k}_\text{b} (T_\text{n} - T_\text{i}) + \text{m}_\text{n} | \vec{V} - \vec{U} |^{2} ]
\end{aligned}
\end{equation}
\begin{equation}\label{eq:energyneutral}
\begin{aligned}
	\frac{\partial \text{k}_\text{b} T_\text{n}}{\partial t} = - \nabla \cdot (\text{k}_\text{b} T_\text{n} \vec{U}) + \frac{1}{3} \text{k}_\text{b} T_\text{n} \nabla \cdot \vec{U} - \frac{2}{3} \frac{\nabla \cdot \vec{q}_\text{n}}{n_\text{n}} \\
	+ \frac{2}{3} \frac{\text{m}_\text{n}}{\text{m}_\text{i}+\text{m}_\text{n}}  \nu_\text{in} [ 3 \text{k}_\text{b} (T_\text{i} - T_\text{n}) + \text{m}_\text{i} | \vec{V} - \vec{U}|^{2} ]
\end{aligned}
\end{equation}

\begin{equation}
	\nabla \times \vec{B} = \mu_0 \vec{J}
\end{equation}

\begin{equation}
	\vec{q} = - \kappa \nabla T
\end{equation}

Here $\text{i}$, $\text{e}$, and $\text{n}$ subscripts denote proton, electron and neutral, respectively. The primitive scalars $\text{m}$, $\rho$, $T$ are particle mass, mass density, and temperature, respectively. $\vec{V}$, $\vec{U}$, and $\vec{B}$ are ion velocity, neutral velocity, and the magnetic-field, respectively, with $\vec{g}$ denoting the solar gravitational acceleration near the solar surface. $\kappa$ is the Spitzer conductivity. $\nu_{ij}$ denotes collision rate of species $i$ to species $j$, and $\eta = \text{m}_\text{e}(\nu_\text{en} + \nu_\text{ei})/\text{e}^2n_\text{e}$ is the ohmic resistivity. We use the following collision rates \citep[see][Appendix A]{Pontieu2001} 
\begin{eqnarray}
&\nu_\text{in} &= 7.4 \times 10^{-11} n_\text{n} T^{1/2} \nonumber\\
&\nu_\text{en} &= 1.95 \times 10^{-10} n_\text{n} T^{1/2} \\
&\nu_\text{ei} &= 3.759 \times 10^{-6} n_\text{e} T_\text{e}^{-3/2} \ln \Lambda \nonumber
\end{eqnarray}
where $\ln \Lambda$ is the Coulomb logarithm taken as 10 for simplicity and $n$ is in cm$^{-3}$ and $T = \frac{T_\text{i} + T_\text{n}}{2}$ in K.

% REVISION
An in-depth derivation of weakly ionized plasma equations of different forms has been given by \citet{Leake2014}. In the case we have chosen for the solar chromosphere and its two-fluid nature, the exchange of heat due to frictional-heating is proportional to the mass of the species, for this reason the electron species is ignored. A detailed derivation of frictional-heating has been given by \citet{Vasyliunas2005}. The thermal-energy equations are not total energy equations so this is the only non-conservative variable used; however, this was chosen to allow for the semi-implicit scheme to implicitly solve collisional terms in regions of the chromosphere where the collision frequency is much larger than the Alf\'ven frequency.

% REMOVED
%The current density terms appearing in Equations~\ref{eq:plasmamomentum} and~\ref{eq:neutralmomentum} are a consequence of treating ionized plasma as the center of mass of the plasma in which ions and electrons move in the opposite directions. The center of mass of the plasma moves roughly with the ions. The origins of this formulation has been given by \citet{Song2005}.

In this work, we do not include an electron pressure in our momentum equation, as we expect that it will only matter in regions of small length scales and large current. We are currently working on adding the electron pressure to the code, and we plan future investigations to study its role and importance.

\subsection{Divergence Cleaning}
In order to maintain a divergence-less solution, the simulation employs a general Lagrange multiplier method \citep{glm}. This introduces a new unknown, the Lagrange multiplier $\Phi$, into the model:
\begin{equation}\label{eq:bglm}
	\frac{\partial \vec{B}}{\partial t} = - \nabla \cdot ( \vec{V}\vec{B} - \vec{B}\vec{V} ) - \nabla \times ( \eta \vec{J}  + \frac{1}{n_\text{e} \text{e}} \vec{J} \times \vec{B} ) + \nabla \Phi
\end{equation}
\begin{equation}\label{eq:glm}
	\frac{\partial \Phi}{\partial t} + c_\text{h}^2 \nabla \cdot \vec{B} = - \frac{c_\text{h}^2}{c_\text{p}^2} \Phi
\end{equation}

The scalars $c_\text{h}$ and $c_\text{p}$ stand for the hyperbolic and parabolic speeds of the divergence errors.

It can be easily shown by taking the divergence of Equation~\ref{eq:bglm} that the introduction of divergence in the magnetic-field is correlated with the multiplier.

\begin{equation}
	\frac{\partial \nabla \cdot \vec{B}}{\partial t} = \nabla^2 \Phi
\end{equation}

The divergence errors are assigned into hyperbolic and parabolic parts, where the errors in divergence advects and dissipates away from where they were created. The hyperbolic speed is chosen to be the fastest speed in the domain, the Alfv\'en speed $c_\text{h} = \max(V_\text{A})$, and the ratio $\alpha_\text{p} = \Delta x \frac{c_\text{h}}{c_\text{p}^2}$ \citep[see][p. 5899, 5911]{Mignone2010}, also known as the damping parameter, is kept at a constant of 0.5.

This method keeps the divergence errors bounded to the order of the grid size in the simulation.

\subsection{Numerical Scheme}

The model's finite-volume method and second-order scheme is the Total Variation Diminishing Monotonic Upwind Scheme for Conservation Laws (\textsf{TVD-MUSCL}) to avoid spurious oscillations and capture shocks that may arise in the solution. It is a cell-centered numerical scheme with extrapolated cell-edge variables. The extrapolated variables are calculated using a linear reconstruction with flux limiters. The nonlinear flux is handled using a high-resolution scheme by \citet{Kurganov2000} of the form:
\begin{equation}
	f^*_{j-1/2} = \frac{1}{2} ([f(u^\text{r}_{j-1/2})+f(u^\text{l}_{j-1/2})] - \lambda_{j-1/2}[u^\text{r}_{j-1/2}-u^\text{l}_{j-1/2}])
\end{equation}
\begin{equation}
	f^*_{j+1/2} = \frac{1}{2} ([f(u^\text{r}_{j+1/2})+f(u^\text{l}_{j+1/2})] - \lambda_{j+1/2}[u^\text{r}_{j+1/2}-u^\text{l}_{j+1/2}])
\end{equation}
where $u$ is the primitive variable, $f$ is the nonlinear conservative flux, $l$ and $r$ are the left and right extrapolated states, $j$ is the index of one of the two dimensions, and $\lambda$ is the largest eigenvalue (characteristic speed) of the corresponding $u$.

The left and right states are calculated using the following linear reconstruction:
\begin{equation}
	u^\text{l}_{j-1/2} = u_{j-1} + \frac{1}{2} \phi(r_{j-1}) (u_j - u_{j-1})
\end{equation}
\begin{equation}
	u^\text{l}_{j+1/2} = u_{j} + \frac{1}{2} \phi(r_{j}) (u_j - u_{j-1})
\end{equation}
\begin{equation}
	u^\text{r}_{j-1/2} = u_{j} + \frac{1}{2} \phi(r_{j}) (u_{j+1} - u_{j})
\end{equation}
\begin{equation}
	u^\text{r}_{j+1/2} = u_{j+1} + \frac{1}{2} \phi(r_{j+1}) (u_{j+1} - u_{j})
\end{equation}
\begin{equation}
	r_j = \frac{u_j - u_{j-1}}{u_{j+1} - u_j}
\end{equation}
\begin{equation}\label{minmod}
	\phi(r) = \max(0, \min(1, r)), \lim_{r \to \infty} \phi(r) = 1
\end{equation}
where $\phi$ is the flux limiter and the simple minmod flux limiter \citep{Roe1986} is chosen for simplicity and speed. The flux limiter ensures the scheme is total-variation-diminishing. With the method above the numerical scheme is second-order when the solution is smooth and first-order upwind when a shock is present.

\subsection{Semi-Implicit Time Integration}

The time integration used for the simulation is the simple semi-implicit Eulerian time step.
\begin{equation}
	u^{n+1} + R_\text{I}(u^{n+1}) \Delta t = u^n + R_\text{E}) \Delta t
\end{equation}
The above equation is split into an implicit left side and explicit right side. Here $R_\text{I}$ are the implicit terms and consists of the collisional and diffusive terms, like collision rate, resistivity and heat flux. The collision time scales are typically many orders of magnitude shorter than Alfv\'en time scales so these terms are dominant. The momentum equations (Equations~\ref{eq:plasmamomentum} and~\ref{eq:neutralmomentum}) have collisional terms for each fluid which introduce off-diagonal terms that must be solved implicitly. While the case studied is the chromospheric network, which has sharp magnetic gradients those off-hand terms are also large and this makes a fully implicit code ill-conditioned. This means a linear system is solved in each time step. $R_\text{E}$ contains the explicit terms, which are the source terms and the numerical fluxes calculated as shown above.

The simulation spans hundreds of thousands of time steps, which is why a simple time integration method is used. Conversely, it can be shown using Von Neumann analysis that the general Lagrange multiplier method used in Equation~\ref{eq:glm} is unstable in other semi-implicit methods.

In order to satisfy Courant-Friedrichs-Lewy (CFL) condition is satisfied by finding the fast wave speed and local bulk flow velocities:
\begin{equation}
	c_\text{fast} = \frac{1}{\sqrt{2}} [c_\text{sound}^2 + c_\text{Alfv\'en}^2 + \sqrt{(c_\text{sound}^2 + c_\text{Alfv\'en}^2)^2 - (2c_\text{sound}c_\text{Alfv\'en})^2 } ]^{1/2}
\end{equation}
Where $c$ are the characteristic speeds. This is calculated for both plasma and neutrals (where neutral's is simply the sound speed). Then with a CFL of 0.1 the time steps become:
\begin{equation}
	\Delta t = 0.1 \frac{\Delta z}{\max{(V+c_\text{fast})}}
\end{equation}
Here the maximum of the local speed and fast wave speed sum for each cell is considered. Although the scheme is stable under a higher CFL, a CFL of 0.1 was chosen for accuracy.

\subsection{Code Architecture}\label{sec:code}

The code is written in \textsf{C++} to make prototyping easy and still have low-level access. The \textsf{Armadillo} library~\citep{arma} is used for matrix building and \textsf{CPU} related code as it makes writing linear algebra code more expressive while still calling \textsf{BLAS} subroutines directly. To make use of \textsf{GPU}s while still being brand-agnostic, the \textsf{GPU} library \textsf{ViennaCL}~\citep{viennacl} is used. The library is written in \textsf{OpenCL} and provides iterative solvers and preconditioners, which the code takes advantage of due to its semi-implicit nature. \textsf{ViennaCL} provides high and low-level access to the \textsf{GPU}. Low-level access is needed to make a custom \textsf{OpenCL} kernel of the flux limiter, and the minmod function shown in~\ref{minmod}. 

Convergence tests show that the results are consistent with larger sizes (which the \textsf{GPU}'s memory can hold). This grid size is chosen because it is the smallest size that does not require a trade-off in terms of the physics captured by the model. 

\section{Code Validation}\label{sec:validation}

To validate the code, two tests have been run to ensure that it is producing accurate results. The first is for the neutral fluid, which does not experience Lorentz forces on its own, but does experience pressure forces. The Sod's shock-tube test is used here to demonstrate the code's shock capturing capabilities. The second is the initial value problem tests done by \citet{Soler2013} to show Alf\'ven-wave propagation in a partially ionized collisional medium. This is to demonstrate the collisions and the electro-magnetic forces working as they are supposed to.

\subsection{Sod's shock-tube}

The shock-tube tests studied by \citet{Sod1978} are often used to tests against shock capturing schemes. The shock capturing scheme used in this code is the \textsf{TVD-MUSCL} scheme which is known to perform well against these tests. The fluid used in this case is the neutral fluid in the absence of collisions. The thermal-energy Equation~\ref{eq:energyneutral} has been modified to a pressure equation similar to Sod's to ensure reproducible results; every other equation has not been changed.

\begin{figure}[ht]
	\includegraphics[width=0.9\textwidth]{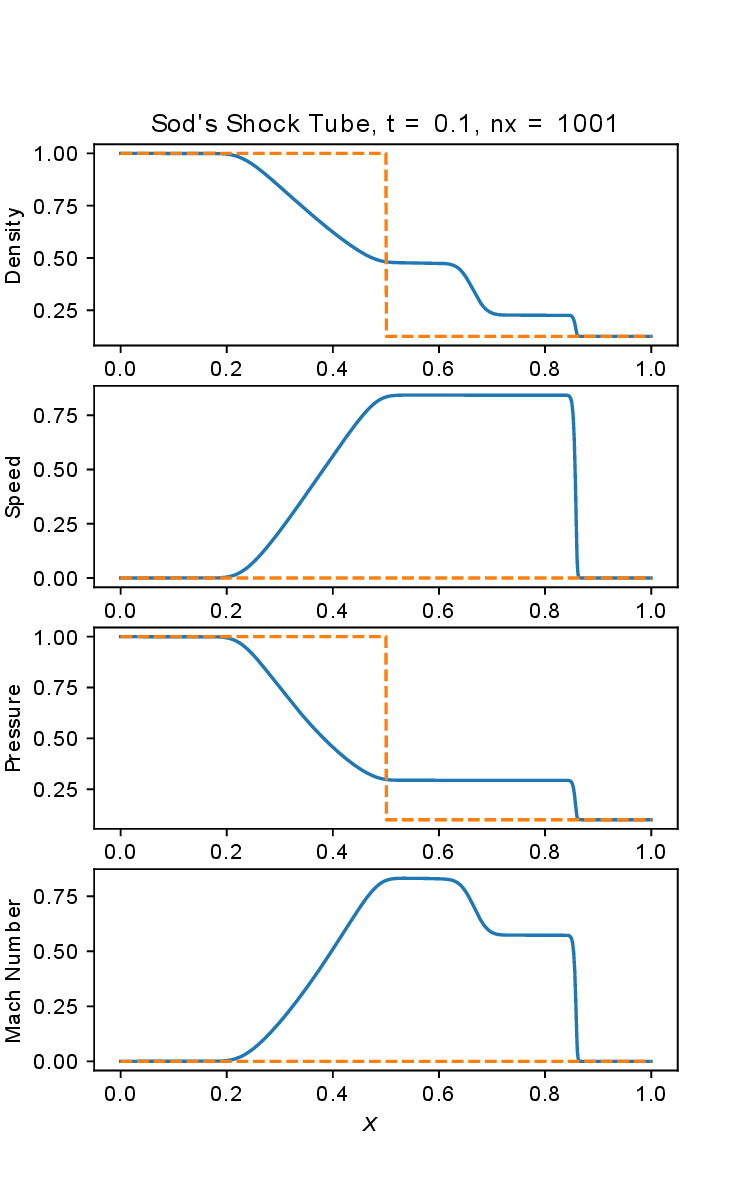}
	\caption{Neutral-fluid shock-tube tests. The shocks are being captured appropriately due to the \textsf{TVD-MUSCL} scheme used. All quantities shown are in normalized units. The \textit{orange dashed lines} are the initial conditions. The \textit{blue solid lines} are the results at time $t=0.1L/\max{(c_\text{sound})}$.}\label{fig:shocktube}
\end{figure}

Figure \ref{fig:shocktube} shows the results, and they mimic the results shown in the article by~\citet{Sod1978}. The runs were done in arbitrary units. The initial conditions are $\rho_\text{l} = 1$, $V_\text{l} = 0$, $p_\text{l} = 1$ and $\rho_\text{r} = 0.125$, $V_\text{r} = 0$, $p_\text{l} = 0.1$. The number of grids chosen were 1001 and the results are taken at time $t = 0.1$. The boundary-conditions were all static Dirichlet boundary-conditions matching the initial conditions for either side. The five regions mentioned in the original article by~\citet{Sod1978} can be clearly seen in the density plot.

\subsection{Initial-Value Problem}

\begin{figure}[ht]
	\includegraphics[width=1\textwidth]{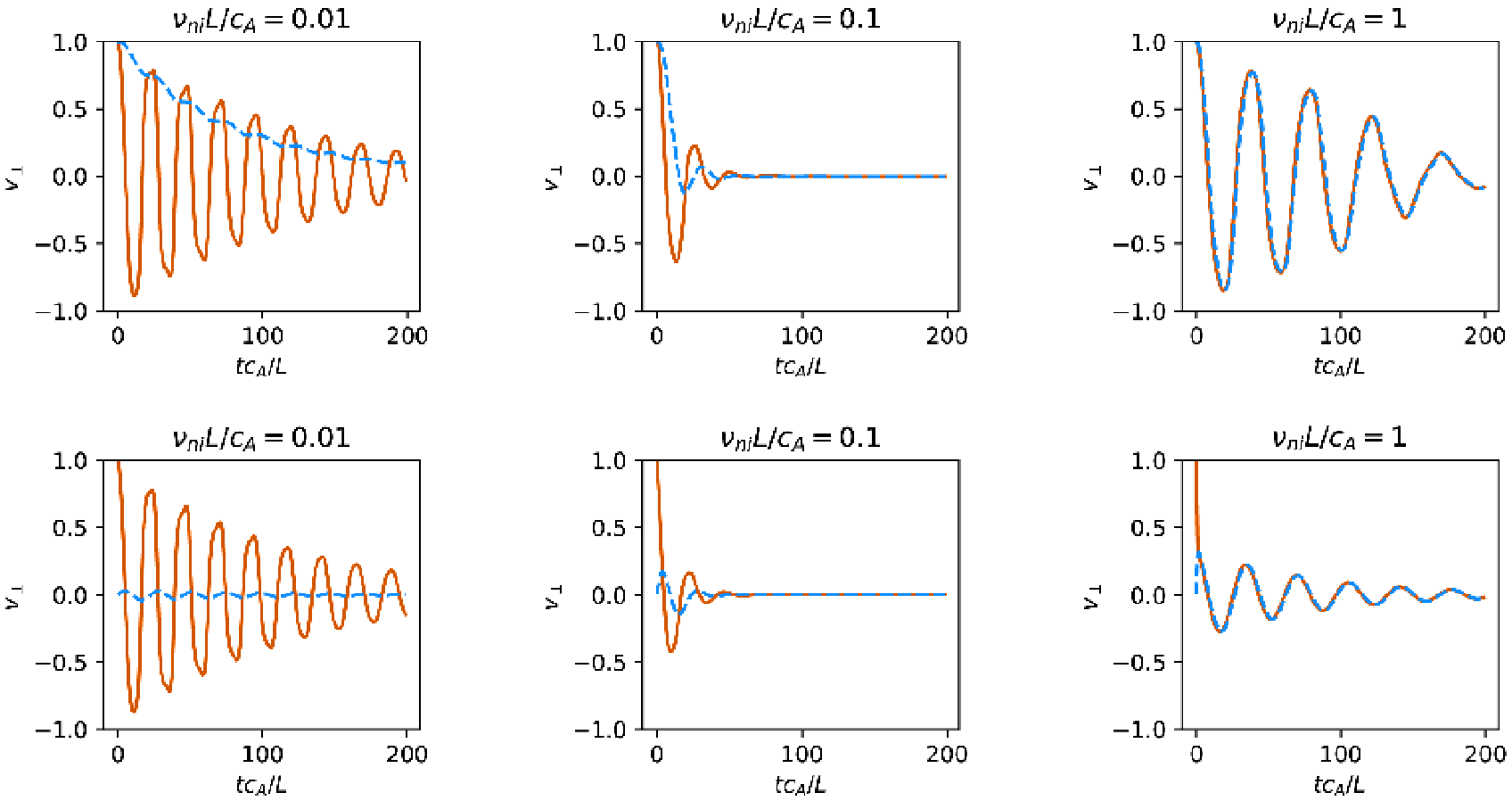}
	\caption{The \citet{Soler2013} initial-value-problem tests. Velocity is taken at $z=0$ of the standing Alf\'ven-wave. \textit{Solid red} is the ion velocity, \textit{dashed blue line} is neutral velocity. The \textit{top row} is where the initial neutral velocity is the same as the ion velocity and the \textit{bottom row} is where the initial neutral velocity is at zero. The results resemble reasonably well the tests of \citet{Soler2013}.}\label{fig:soler}
\end{figure}

\citet{Soler2013} ran many tests of Alfv\'en-wave propagation in a partially ionized medium. The time-dependent test that was presented was an initial-value-problem where the initial value of the velocity is $v_{\perp}(z) = \cos{(20\pi z / L)}$ in the region of $z = [-10L, 10L]$ corresponding to a fundamental mode of a standing Alfv\'en-wave. Three tests are shown of two different cases, where the initial value of the neutral velocity is similar or zero. The three tests are three different orders of magnitude of the collision rates. It is important to note the \textsf{MolMHD} code used in this article is a fundamentally different set of equations, using the primitive variables instead of the conservative ones. However, despite these differences, the results shown in Figure~\ref{fig:soler} are similar overall.

\section{The Simulation Domain}\label{sec:domain}

In the horizontal dimension, we simulate a supergranule similar to that described by \citet{Song2017}. Because of symmetry, our simulation domain contains half of a supergranule, from the center of a network with strong-field to the center of internetwork at the photospheric boundary. If we further assume that the neighboring supergranules have the same polarity as the one in the simulation domain, there is no reconnection between the supergranules of interest.  In this case the normal component of the field [$B_x$] is zero at the left and right boundaries. The right and left boundaries act as mirrors to reflect any normal component of the field. Similarly, the normal component of the velocity is also set to zero at the right and left boundaries, \textit{i.e.} there is no horizontal flow from one supergranule to another at the chromospheric altitude although inter-supergranule flow is possible at coronal heights and in or below the photosphere. 

Since we only simulate the chromosphere and treat the transition region above and photosphere below as upper and lower boundaries, respectively, the photosphere below is made up as a Dirichlet boundary-conditions. Below the surface of the photosphere, the Sun is highly dynamic. The assumption made in this model is that the Alfv\'en waves are generated at the photosphere and propagate while undergoing damping towards the transition region. For this reason, to imitate the energy flux and activity of the photosphere, a driver at the bottom of the simulation domain is made as a Dirichlet boundary-conditions. Using the broadband-spectrum-driving method of \citet{Tu2013}, the perpendicular velocity is a continuous oscillation with a broadband time series of the form:
\begin{equation}
\begin{aligned}
	V_{\perp} (t, z=0) = V_\text{b} [ \sum_{k=0}^s {(\omega_k/\omega_\text{p})}^{5/6} \cos{(\omega_k t + \Phi_k)} \\
	+ \sum_{j=s+1}^m {(\omega_\text{p}/\omega_j)}^{5/6} \cos{(\omega_j t + \Phi_j)} ],
\end{aligned}
\end{equation}
where $V_{\perp}$ will be the imposed speed on the lower boundary in the direction normal to the $x$--$z$ plane. $V_\text{b}$ is the amplitude of the broadband spectrum, 2\,km\,s$^{-1}$, to ensure an appropriate photospheric flux of $2 \times 10^7$\,erg\,cm$^{-2}$\,s$^{-1}$ \citep[see][section 3.2]{Tu2013}, $\omega_j$ and $\omega_k$ are angular frequencies, $\omega_\text{p}$ is the peak angular frequency at 1.65\,mHz. The driver also has random phases $\Phi_j, \Phi_k$ each of which is a multiple of 2$\pi \times$ a random number between 0 and $s+m$. Here $s$ is the number of frequencies chosen smaller than the peak frequency and $m$ is the number of frequencies larger than the peak frequency. For the case of this simulation $s=100$ and $m=500$. 

\begin{figure}[ht]
		\centering
		\includegraphics[width=1\textwidth]{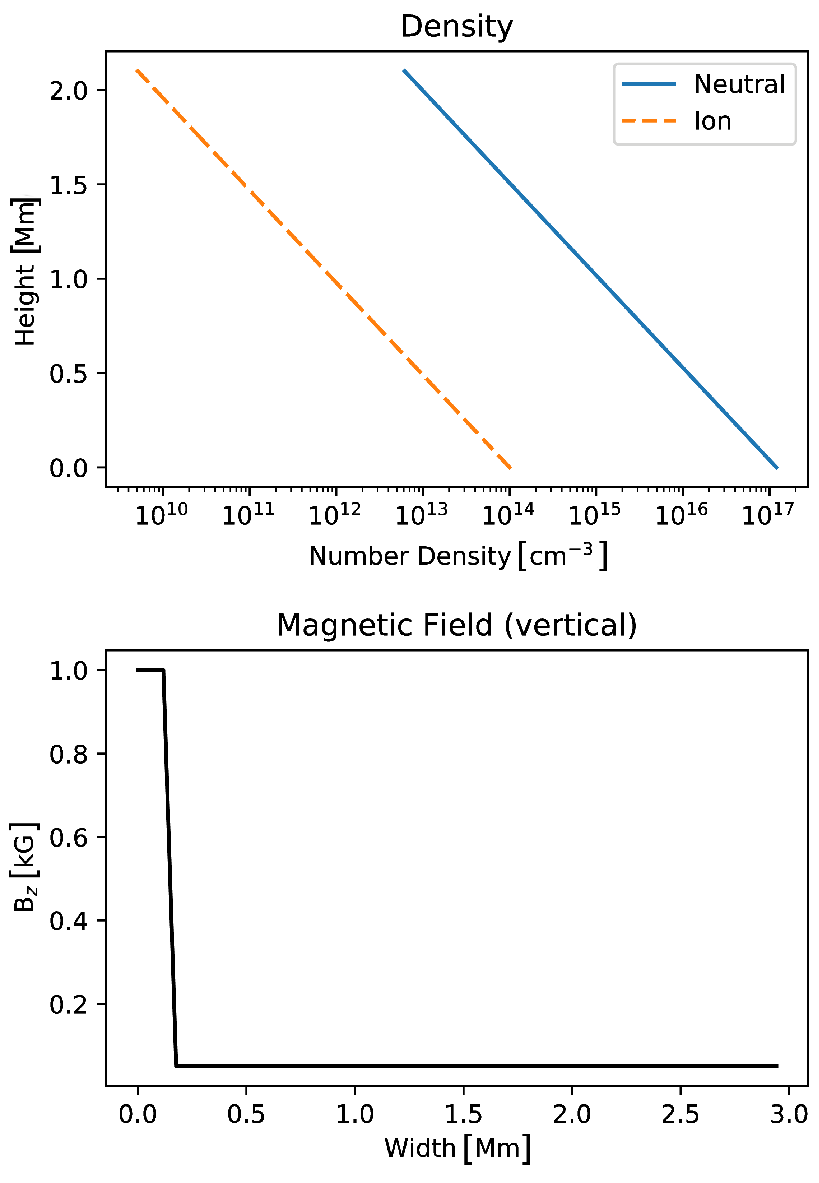}
		\caption{The initial conditions for the current simulation. Ions and neutral density are initial conditions are stratified and hydrostatic. The \textit{blue line} is the neutral number-density and the \textit{orange line} is the ion number-density. The horizontal magnetic-field varies horizontally but not vertically.}\label{fig:ic}
\end{figure}

The number-density of the ions and neutrals are specified at the boundary as the photospheric values from \citet{AvrettLoeser2008} of 10$^{20}$ m$^{-3}$ and 10$^{23}$ m$^{-3}$, respectively. As an initial condition, a stratified hydrostatic profile is made for the ion and neutral densities $n(z) = n(z=0) e^{-\text{m}_\text{i}\text{g}z/\text{k}_bT_0}$; this can be seen in Figure~\ref{fig:ic}. Here $T_0$ = 7000\,K which is the temperature imposed on the photosphere and also the initial condition for both ionized plasma and neutrals in the entire domain. An isothermal approach was chosen initially because the chromosphere's temperature does not change much, between 6000\,K to 8000\,K, between the photosphere and the transition region. At later times, the temperatures are determined from the energy Equations~\ref{eq:energyion} and~\ref{eq:energyneutral}. 

A requirement in some MHD codes with zero magnetic-field divergence constraints is that the initial condition must also be divergence free. For this reason $B_x(t=0)$ and $B_\perp(t=0)$ is chosen to be zero while $B_z(x, t=0)$ varies in the horizontal $x$-direction. It consists of regions of strong-field, $\approx$1\,kG, for the network region and weak-field, $\approx$50\,G, for the internetwork region with a sharp linear drop in between. The 1\,kG region only takes up 0.05\,Mm in the horizontal direction and is in the left part of the domain, while the weak region part is in the rest of the domain. The goal is to allow this initial condition to quickly relax to a canopy because of the large $\beta$ change due to height without solving a force-free field. At the bottom boundary,  a Dirichlet condition is applied to $B_z(x)$ but for the other two components the field is set to Neumann conditions, $\frac{\partial B_{z,\perp}}{\partial z}\vline_{z=0\,\text{km}} = 0$. This configuration resembles the stem of the canopy and the magnitude of the supergranule strengths at the photosphere. The bottom boundary's $B_z(x)$ is an unchanging Dirichlet condition to mimic the convection cell underneath. This is also shown in Figure~\ref{fig:ic}.

At the top boundary which represents the transition region and the corona above it, the model uses Neumann conditions to allow free-flow of the plasma through it (\textit{i.e.} $\frac{\partial \textbf{u}}{\partial z}\vline_{z=2100\,\text{km}} = 0$).

In the simulations and results shown in this article the domain is 15\,Mm horizontally (half the size of the magnetic internetwork) and 2.1\,Mm vertically (typical thickness of the chromosphere). It is a uniform Cartesian grid of 256$\times$128 where $\Delta x = 58.6$\,km and $\Delta z = 16.4$\,km. This is the largest cell size chosen for speed and memory considering the usage of a dedicated scientific computing \textsf{GPU} (see Section~\ref{sec:code}). This is larger than the length scale of the collisionally reduced Alfv\'en-wavelength ($\frac{V_\text{A}}{\nu_\text{in}}$) which means for this simulation the plasma and neutrals are coupled. Since this is a 2.5D simulation all spatial derivatives in the perpendicular direction are taken as zero ($\frac{\partial}{\partial \perp} = 0$).

All momentum components are set initially as zero.

%%%%%%%%%%%%%%%%%%%%%%%%%
%  Results & discussion
%%%%%%%%%%%%%%%%%%%%%%%%%%

\section{Results}\label{sec:results}

\begin{figure}[ht]
	\includegraphics[width=1\textwidth]{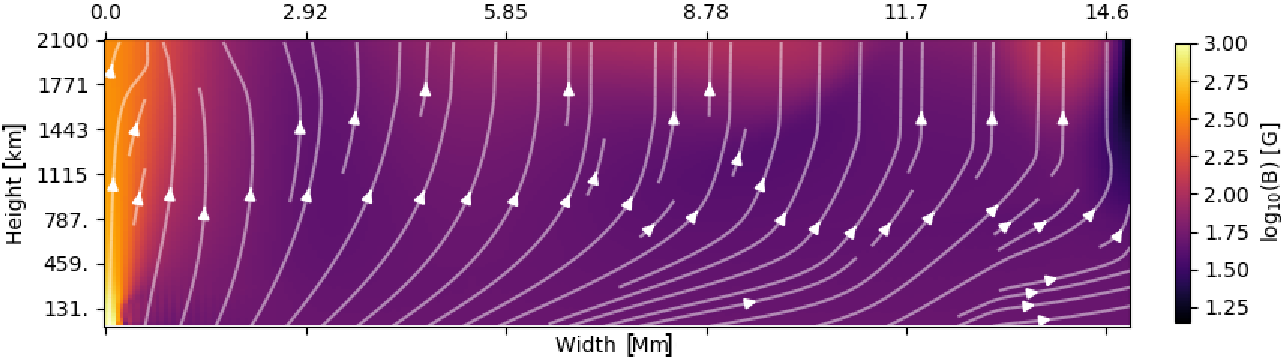}
	\caption{Formation of a canopy seen after around 348\,ms (13000 time steps). The initial condition of a concentrated strong-field region (1\,kG) and weak-field region (5G) creates vertical asymmetry as plasma-$\beta$ drastically decreases. The left and right boundaries are mirrored so effects of other granules behaving the same way are considered. Because of the short time period this is \textit{not} a relaxed system but simply a consequence of the pressure imbalance frequently seen in the chromosphere.}\label{fig:granulegeometry}
\end{figure}

Figure~\ref{fig:granulegeometry} shows a snapshot on the chromospheric solution after 348\,ms. It shows that our simulation produces a magnetic-field topology that is characterized by a canopy or ``wine glass" shape \citep{Rutten2007}, with a strong-field of about 1\,kG at the stem on the left, and a weaker field of about 50\,G near the upper boundary of the simulation domain, near the transition region.  This topology is obtained fairly quickly, in around $\approx$350\,ms, because of the magnetic pressure imbalance imposed in the initial conditions. This is not a relaxed system, as the canopy holds for a few seconds before an eruption starts forming. 

\begin{figure}[ht]
		\centering
		\includegraphics[width=0.9\textwidth]{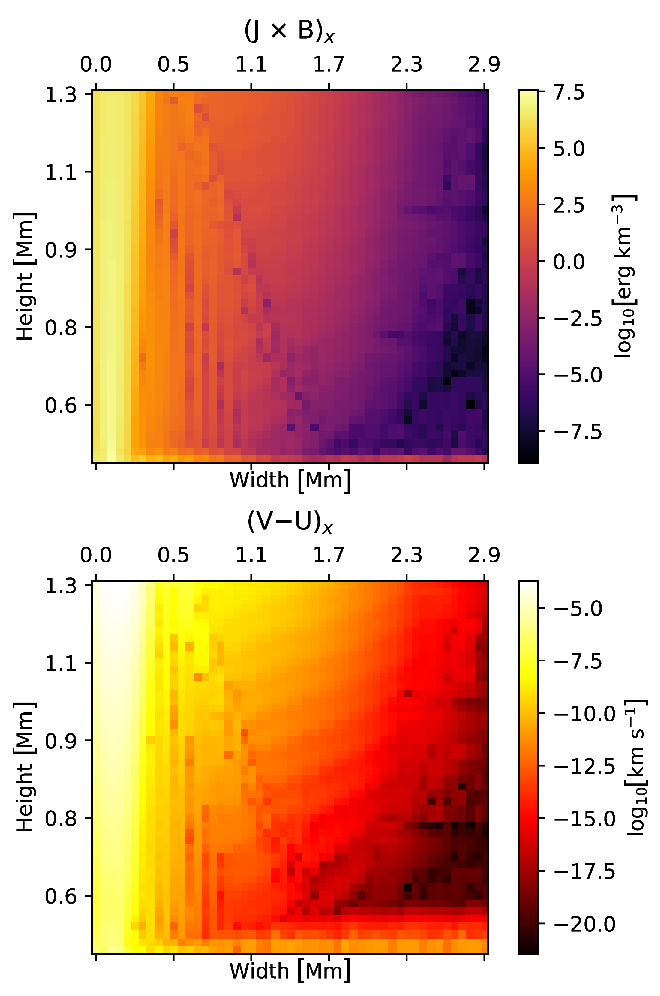}
		\caption{Lorentz force and difference in species' speeds in the horizontal direction at $t=101$\,ms--right before before the jet forms. The difference in speeds cause frictional-heating which causes enhanced heating in the strong-field region. This leads to the vertical thermal-pressure imbalance that causes the upwards motion of the jet.}\label{fig:causality}
\end{figure}

Figure~\ref{fig:causality} shows the force that is leading to the jet forces. It is a snapshot of $t=101$\,ms, which is after the simulation start and before the jet formation.

\begin{figure}[ht]
		\centering
		\includegraphics[width=0.75\textwidth]{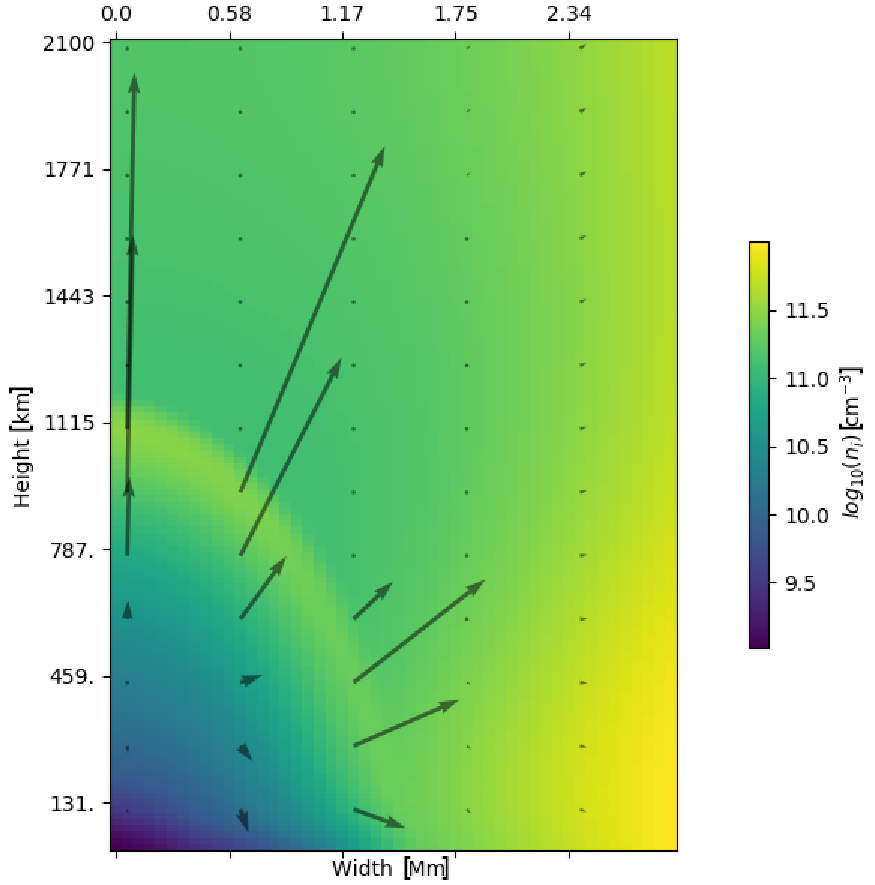}
		\caption{An enlargement into the shock wave front of the jet. The arrows represents the direction and magnitude of ion flow. The color bar represents the neutral number-density in cm$^{-3}$}\label{fig:spiculeshock}
\end{figure}

Figure~\ref{fig:spiculeshock} shows a formation of what resembles a jet--a parcel of plasma quickly moving upward towards the transition region. The shock spontaneously forms after around 20\,{seconds} and the jet speeds up and shoots up outside of the chromosphere into the corona in about five seconds. 

\begin{figure}[ht]
    \centering
    \includegraphics[width=1\textwidth]{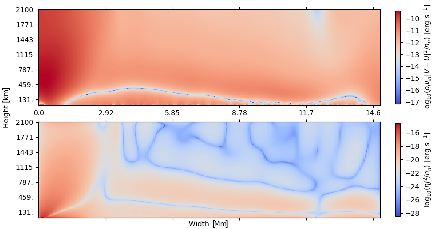}
    \caption{Heating of the shock origin before the formation of the jet at $t=18.9$\,{seconds}. \textit{Top} image is the frictional-heating per neutral particle. \textit{Bottom} image is the ohmic-heating per neutral particle. The frictional-heating is orders of magnitude larger than the heating due to currents. The sharp white and blue regions is where $|V-U|=0$ and calculating the logarithm of a quantity that tends to zero can show up as numerical artifacts.}\label{fig:prespicule}
\end{figure}

In Figure~\ref{fig:prespicule}, we show the distribution of the two heating terms--the ohmic-heating [$\eta J^2$] and the frictional-heating [$\nu_\text{in} \rho_\text{i} |\vec{V} - \vec{U}|^2$] in the simulation domain, right before the eruption. The heating rates are normalized per neutral particle. It can be seen that the frictional-heating dominates the ohmic-heating by orders of magnitude, especially in the strong-field region where the shock forms.

\begin{figure}[ht]
	\centering
    \includegraphics[width=1\textwidth]{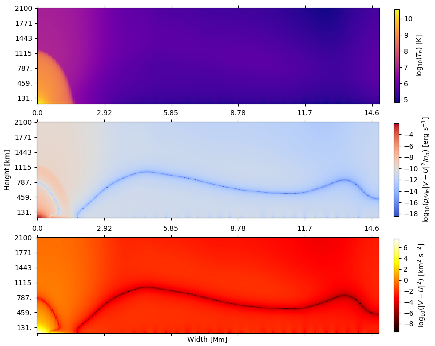}
	\caption{The jet in the middle of its journey to the corona at $t=26.8$\,{seconds}. From \textit{top to bottom}, the temperature of the neutral plasma, the frictional-heating per neutral particle, and the difference in speeds $\| \vec{V} - \vec{U} \|^2$ of the charged and neutral plasma. The shock can be clearly seen in the bottom-left corner of the simulation where the magnetic flux concentration is strongest. Due to large plasma beta a ``ring" of slow moving plasma is formed and persists in the shock allowing the shock to start forming. The sharp blue and black regions are due to $|V-U|=0$, where calculating the logarithm of a quantity that tends to zero creates numerical artifacts.}\label{fig:midspicule}
\end{figure}

Figure~\ref{fig:midspicule} shows the neutral temperature, the frictional-heating, and the velocity difference [$| \vec{V} - \vec{U}|^2$] during the emergence of the jet to the corona (at $t=26.8$\,{seconds}). The plots show the formation of an appreciable shock, which separates a low number-density, high temperature, and strong heating region from a high number-density, lower temperature, and weaker heating in the rest of the medium. The figure also shows a very clear difference in heating rates on the two sides of the shock, where the front of the shock travels at super sonic speeds. The shock propagates supersonically until the plasma is ejected when the chromosphere is depleted and starts to fill up again with flows going back down. Downstream of the shock, the plasma flow is returning due to a cavity left behind, where upstream of the shock, the chromospheric plasma is still unchanged.

\begin{figure}[ht]
	\centering
    \includegraphics[width=1\textwidth]{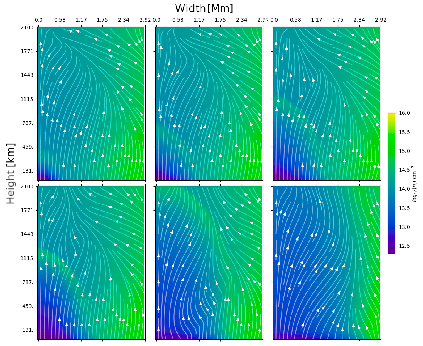}
    \caption{Time series of the shock. From \textit{left to right and top to bottom} the times are $t=26.751$\,{seconds}, 26.785\,{seconds}, 26.831\,{seconds}, 26.842\,{seconds}, 26.844\,{seconds}, 26.849\,{seconds}. The \textit{white streamlines} denote the magnetic-field lines and the number-density of the neutrals are in cm$^{-3}$. This shows the magnetic-field geometry changing due to the jet and not the other way around. The jet is not generated because of the magnetic-field. The last panel shows a cavity left behind due to the jet.}\label{fig:spiculetimeseries}
\end{figure}

\begin{figure}[ht]
	\centering
    \includegraphics[width=1\textwidth]{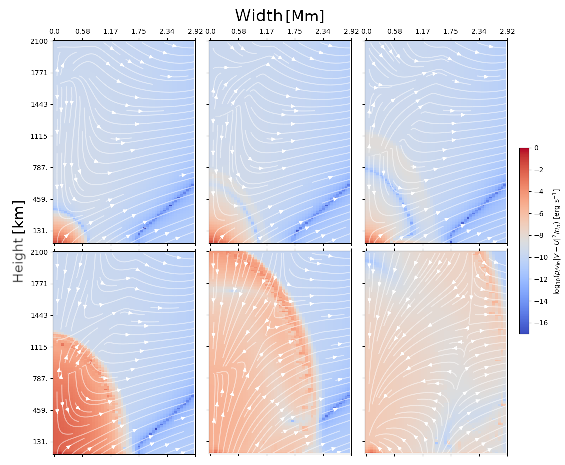}
    \caption{Time series of frictional-heating in the jet. From \textit{left to right and top to bottom} the times are $t=26.751$\,{seconds}, 26.785\,{seconds}, 26.831\,{seconds}, 26.842\,{seconds}, 26.844\,{seconds}, 26.849\,{seconds}. The \textit{white streamlines} denote flow.}\label{fig:heatingtimeseries}
\end{figure}

Figure~\ref{fig:spiculetimeseries} shows a time series of the evolution of the shock number-density and magnetic-field lines. It can be seen that the field lines follow the shock evolution, which means magnetic-field topology (\textit{i.e.} reconnection region) is not a driver of the jet. Figure~\ref{fig:heatingtimeseries} shows the same time series but with contours of frictional-heating and flow streamlines. It clearly shows that the jet is driven by frictional-heating.

\begin{figure}[ht]
	\centering
    \includegraphics[width=1\textwidth]{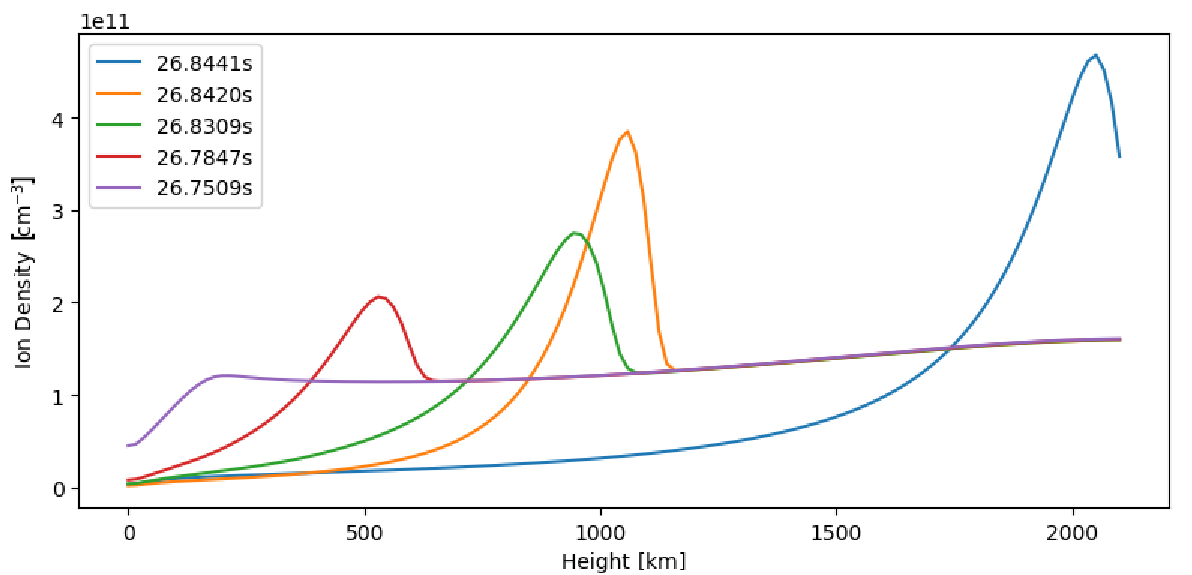}
    \caption{Ion number-density snapshots of shock wave front at line of extraction $x=0.6$\,Mm. Ion number-density is in units of $10^{11}$\,cm$^{-3}$. The wavefront can be clearly seen rising into the chromosphere.}\label{fig:shockwavefront1d}
\end{figure}

\begin{figure}[ht]
	\centering
    \includegraphics[width=1\textwidth]{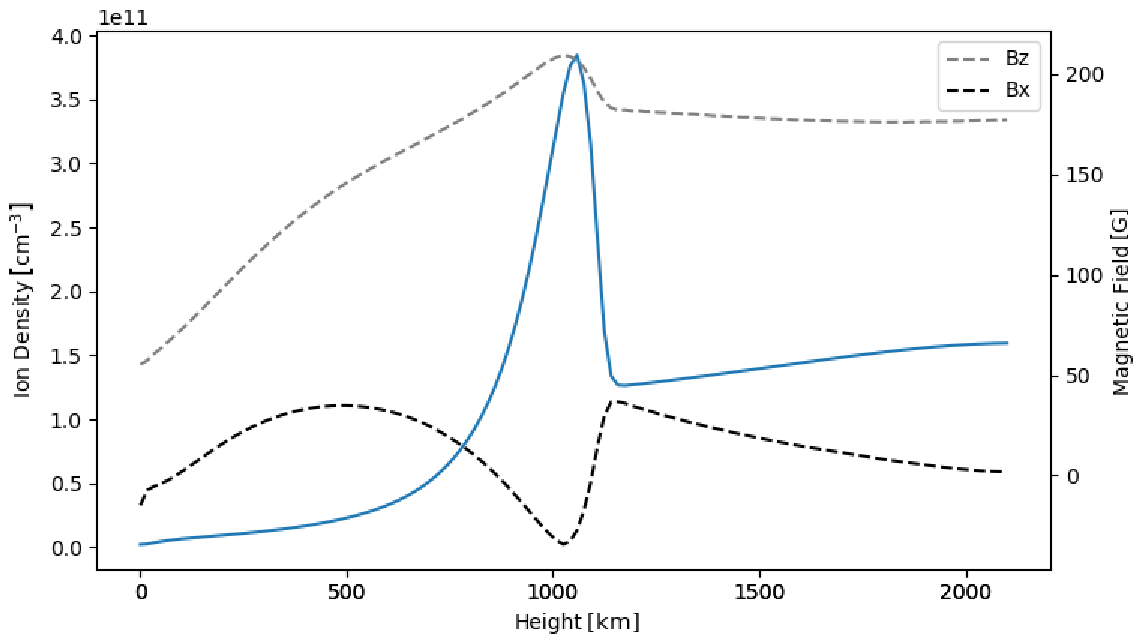}
    \caption{\textit{Blue line} is ion number-density and the \textit{black and gray dashed lines} are $B_x$ and $B_z$ respectively at $t=26.83$\,{seconds} and $x=0.6$\,Mm. Ion number-density is in units of $10^{11}$\,cm$^{-3}$. The magnetic-field is changed behind the shock wave front.}\label{fig:spiculefield1d}
\end{figure}

\begin{figure}[ht]
	\centering
    \includegraphics[width=1\textwidth]{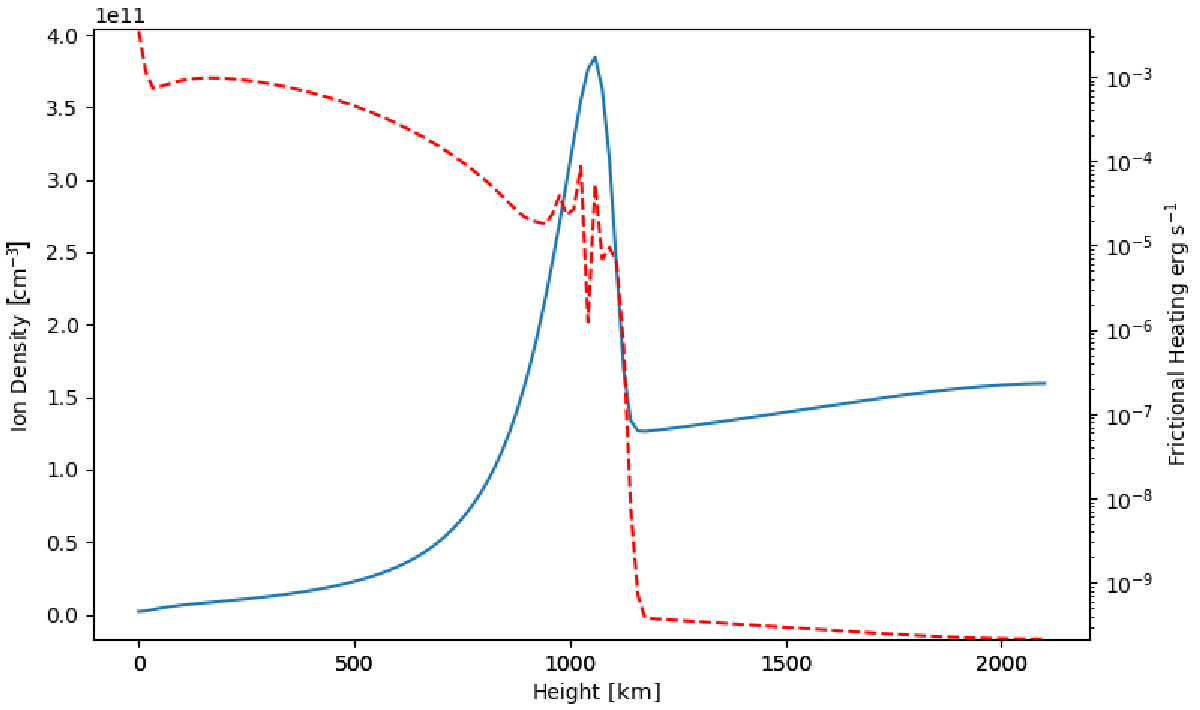}
    \caption{\textit{Blue line} is ion number-density and \textit{red dashed line} is frictional-heating per neutral particle at $t=26.83$\,{seconds} and $x=0.6$\,Mm. Ion number-density is in units of $10^{11}$\,cm$^{-3}$. The frictional-heating per neutral particle behind the wavefront is orders of magnitude higher than in front of it.}\label{fig:spiculefriction1d}
\end{figure}

\begin{figure}[ht]
	\centering
    \includegraphics[width=1\textwidth]{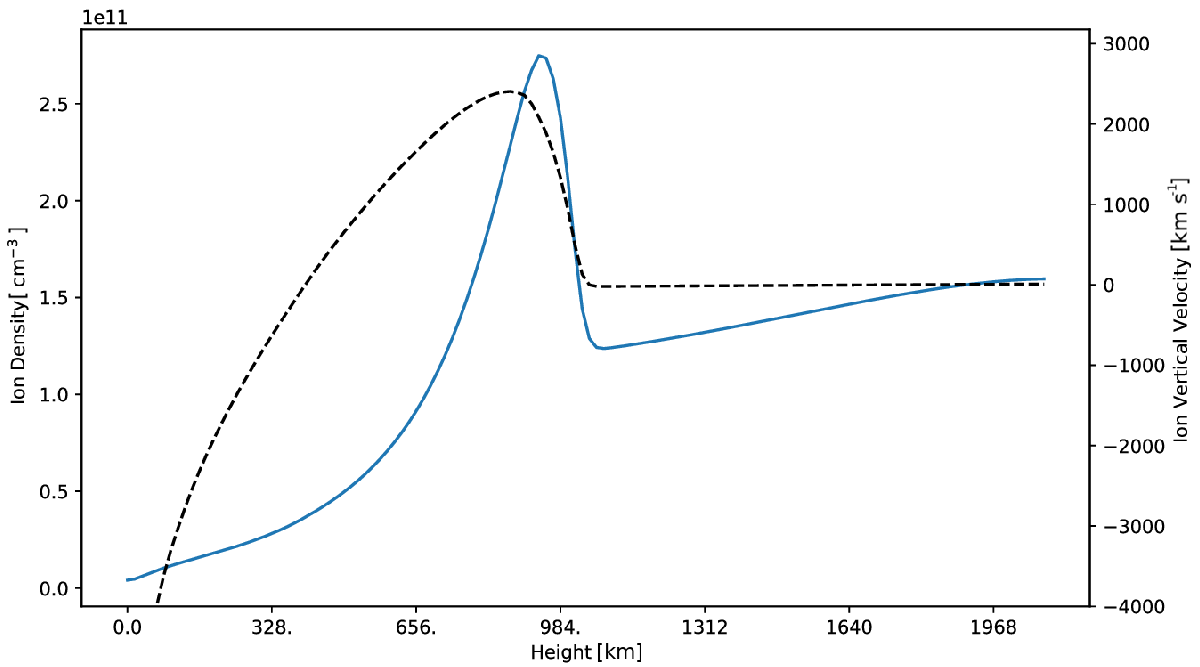}
    \caption{\textit{Blue line} is ion number-density and \textit{black dashed line} is vertical ion velocity at $t=26.83$\,{seconds} and $x=0.6$\,Mm. Ion number-density is in units of $10^{11}$\,cm$^{-3}$. The jet speed is supersonic.}\label{fig:spiculespeed1d}
\end{figure}

Figures~\ref{fig:shockwavefront1d},~\ref{fig:spiculefield1d},~\ref{fig:spiculefriction1d}, and~\ref{fig:spiculespeed1d} show one-dimensional extractions of the solution during the shock emergence along a vertical column at $x=0.6$\,Mm.  These plots exemplify the cause and effect of the shock. The first plot shows the time series of the wavefront of the shock climbing in the chromosphere. The second and third show that the frictional-heating jump follows the shock, and that the magnetic-field changes correspond to the jet behind the shock. Figure~\ref{fig:spiculespeed1d} shows a velocity profile of the jet.

\subsection{Discussion}

A magnetic canopy forms due to the magnetic and thermal-pressure imbalance of the strong-field regions from the photosphere and the weak-field but high-temperature regions scaling up to the transition region. This is consistent with a two-fluid analytic model that was presented by \citet{Song2017}. It shows that the model can provide numerical solutions for chromospheric conditions. However, the magnetic canopy does not relax or reach a steady state due to the nature of the collisional model and not in the timescales of the current simulation. The supergranule canopy may not last because the photospheric power flux, generated at the lower boundary, continuously adds energy into the system. Small scale processes or highly dynamic events like jets override the canopy magnetic-field because of large plasma speeds. Our results show that the jet has immediate effects on the magnetic-field due to induction and not \textit{vice versa}.

The largest Alfv\'en speeds seen in the simulation is 11\,Mm\,s$^{-1}$ which means that in the 30\,{seconds} of simulation, magnetic effects can travel through the domain at least three times with constant density and even less in a hydrostatic density. Alfv\'en waves generated as the drivers for frictional-heating at the bottom boundary do not reflect at the top boundary enough times to warrant a time-averaged study of the entire heating of the chromosphere, nor does it have enough energy to drive the jet of the same kind as presented. The presence of the shock also inhibits any long-term ideal studies of the atmosphere since it depletes and cavities form with the jet. In a future study, we will see how Alf\'ven waves affect the chromosphere with force-free initial conditions so as to not form any jets that can destroy a canopy configuration.

Although the frictional-heating is large in the entire magnetic network column, an instance of no frictional-heating creates a separation of upstream and downstream fluids preceding the shock. This can be seen in Figure~\ref{fig:prespicule} and~\ref{fig:midspicule} and is due to a high temperatures and density, which causes the two fluids to move in unison and therefore $\rho_\text{in} \nu_\text{in} \|V-U\| = 0$. The cause of the frictional-heating is not Alfv\'en-wave damping but the imbalanced initial conditions and the horizontal $\vec{J} \times \vec{B}$ force. The lower medium heats faster than above it, and this creates the shock in the lower atmosphere that then propagates due to thermal-pressure.

It has been suggested that the driver of these kinds of spicule-like events is magnetic reconnection that originates from the vicinity of a large magnetic flux concentration in the network \citep{Pontieu2007}. In our simulation, the frictional-heating heats a fluid downstream of the unity speed ring larger than that of upstream causing a shock wave as expected from a Type I spicule. However the simulation does not show reconnection effects with enough energy to drive a spicule. Although MHD simulations do not mimic reconnection directly, currents present in reconnection events would accelerate electrons into a collision due to the dense and hot medium surrounding it. Therefore, magnetic reconnection is directly related to ohmic-heating in the chromosphere. If comparing the dominating heating terms, frictional-heating dominates ohmic-heating, which means {\it collisions are causing these jets to form and not reconnection-like currents}, then the currents are not due to anti-parallel components but a gradient of a magnetic-field line towards the same direction. The size and extreme temperatures of the jet in the simulation closely resemble chromospheric jets~\citep{Shimojo2000,Nishizuka2011}.

In Figure~\ref{fig:shockwavefront1d} the jet can be seen accumulating a shock of number-density $10^{11}$\,cm$^{-3}$. This density is consistent with IRIS observations of spicules in the corona~\citep{Alissandrakis2018}. The simulation does show that changes in the magnetic-field happen behind the shock wave front's direction of motion. Figure~\ref{fig:spiculefield1d} implies the change in the magnetic-field is a consequence of the jet and the rising magnetic-field density in the plane of the simulation is due to compression from the shock itself. The shock in the frictional-heating term [$\rho_\text{i} \nu_\text{in} \| \vec{V} - \vec{U} \|^2/n_\text{n}$] seen in Figure~\ref{fig:spiculefriction1d} coincides exactly with the shock showing enhanced heating behind the shock front. The cause of the shock traveling upward is the thermal-pressure behind and in front of the shock.

The jet's temperatures and speed are much larger than observed on the Sun. This is mostly due to the fact that in our model we have not included the radiative loss in the energy equation. The larger-than-reality shock heating because of the greater flow speed may also further amplify the temperature. The over-estimated temperature of the downstream medium may also lead to larger collision frequencies. This growing downstream temperature has a compounding effect on the thermal-pressure gradient downstream and upstream and causes the speeds to go supersonic. Further studies will incorporate the photo-chemistry and also radiative losses. These implementations will help to see how they affect shock formation and the energy balance inside the chromosphere. It is also important to note that this shock travels at super-Alfv\'enic speeds.

\citet{KuzmaApJ2017} carried out a numerical simulation of a solar spicule by imposing a vertical pulse at the solar limb. In contrast, the shock shown in our simulation self-consistently evolves without requiring a push on the plasma, and the jet is a consequence of thermal effects. Thus, it provides a complete picture of the chromospheric-heating conditions for shock formation. The spicule speeds in \citet{KuzmaApJ2017} match more closely to the observed 25\,km\,s$^{-1}$ than in our simulation, but that is to be expected in the case where the pulse is controlled and the simulation is single-fluid. \citet{KuzmaApJ2017} also investigated adiabatic and non-adiabatic studies of spicules and the wave structures found in the plasma. The jets in our simulation are different from the imposed pulses of~\citet{Kuzma2017} since that is a cold dense plasma that has been imposed, while in our simulation, it is a hot large plasma jet that evolved self-consistently.

\citet{Song2014} found that by subtracting the two fluids' momentum equations the time evolution of the speed differences [$\text{d/dt}|\vec{V}-\vec{U}|$] which leads to frictional-heating, is caused and maintained by magnetic pressure and thermal-pressure differences between the fluids. This is the cause for the heating near the photosphere as shown in Figure~\ref{fig:causality}. The Lorentz force in the horizontal direction dominates other forces and causes a difference in the species' horizontal speeds. This leads to the frictional-heating that creates a vertical thermal-pressure difference that causes the jet's upwards motion. \citet{Tu2013}, who studied a time-averaged heating in the chromosphere, found that ohmic-heating dominated in the lower chromosphere and frictional-heating dominated in the upper chromosphere, which adequately explained the heating of the chromosphere. This picture seems to be different from the results of our simulation, as frictional-heating dominated ohmic-heating in the lower chromosphere, but this is most likely due to the field strength used in the two studies. Their study limited the field to less than 100\,G, but our model includes a region of 1\,kG field. The one order of magnitude higher field decreases the ohmic-to-frictional-heating ratio by two orders of magnitude.

We use our model to demonstrate the importance of collisions in the chromosphere. A model without collisions is incomplete as neutral particles dominate the charged particles in this small layer of the Sun. By simply adding collisional terms we show that the simulation replicates chromospheric conditions well. Such a model can be used as an interface between the (fully) neutral photosphere and the (fully) ionized corona. \citet{Hillier2016} investigated 1D slow-mode shocks in chromospheric conditions and how magnetic energy can be converted to frictional-heating in these cases.

We plan to dedicate a future study to investigate how the initial and boundary conditions affect the frictional-heating and the formation of the eruptive feature. We would like to apply periodic boundary-conditions and turn off the Alfv\'en-wave driver, while shortening the domain to just the lower chromosphere to study the formation more carefully and reduce boundary-conditions effects.

Finally, this work serves as a step towards a better understanding of the solar chromosphere. The {\it Parker Solar Probe} mission \citep{SolarProbe16} is expected to shed light on the dominant processes in the solar atmosphere, and these processes will be implemented in the model. 

\section{Conclusion}\label{sec:conclusion}

In this article, we present a self-consistent, two-fluid MHD model to study the chromosphere. The model includes ion--neutral interactions, which allows us to investigate the role of frictional-heating in the dynamics of the chromosphere.  The very high collision rates in the partially ionized chromosphere stabilizes the code so that no numerical viscosity is needed.  

The model obtains preliminary results which require further investigation. Our simulation suggests that, between network and internetwork regions, a jet could potentially be driven purely by thermal processes, especially the frictional-heating, and not by magnetic reconnection. Thus, it is crucial to include the neutrals in any attempt to model the lower chromosphere.

The discrepancy between observational speeds and speeds in the simulation could be explained by the need to include the effects of photo chemical reactions in the chromosphere and the radiative losses. Further refinement of the model with photo-chemical effects and radiation.  

%%%%%%%%%%%%%%%%%%%%%%
% Acknowledgments
%%%%%%%%%%%%%%%%%%%%%%

\section*{Acknowledgments}

We thank an unknown referee for their useful comments and suggestions. O. Cohen was supported by NASA Living with a Star grant number NNX16AC11G. J. Tu and P. Song were supported by NSF grants AGS-1702134 and AGS-1559717.

%%%%%%%%%%%%%%%%%%%%%%%
% Bibliography
%%%%%%%%%%%%%%%%%%%%%%%
\bibliographystyle{spr-mp-sola}
\bibliography{SOLA_18_244R3}
%%%%%%%%%%%%%%%%%%%%%%%
% Figures
%%%%%%%%%%%%%%%%%%%%%%%

\end{article}
\end{document}